\documentclass[11pt] {article}
\usepackage{graphicx}       % for eps figures
\usepackage{bm}         % for bold math symbols
\usepackage{amsfonts}
\usepackage{amsmath}        % for \text and such
\usepackage{latexsym}

\begin{document}
\title{Interference of composite bosons}
\author{Thomas Brougham$^1$$^{\ast}$$\thanks{$^\ast$ Corresponding author.  Email: thomas.brougham@gmail.com}$, Stephen M. Barnett$^{1,2}$ and Igor Jex$^1$\\
$^1$ Department of Physics, FNSPE, Czech Technical University in Prague, B\v{r}ehov\'a 7,\\ 115 19 Praha 1, Czech Republic;\\ $^2$ SUPA, Department of Physics, University of Strathclyde, Glasgow G4 0NG, United Kingdom}
%\address{$^1$ Department of Physics, FNSPE, Czech Technical University in Prague, B\v{r}ehov\'a 7, 115 19 Praha 1, Czech Repbulic; $^2$ SUPA, Department of Physics, University of Strathclyde, Glasgow G4 0NG, United Kingdom}
\maketitle
\abstract{We investigate multi-boson interference.  A Hamiltonian is presented that treats pairs of bosons as a single composite boson.  This Hamiltonian allows two pairs of bosons to interact as if they were two single composite bosons.  We show that this leads to the composite bosons exhibiting novel interference effects such as Hong-Ou-Mandel interference. We then investigate generalizations of the formalism to the case of interference between two general composite bosons.  Finally, we show how one can realize interference between composite bosons in the two atom Dicke model.
\\keywords: interference of bosons, quantum optics, Hong-Ou-Mandel effect.}

\section{Introduction}
\label{secI}
One area in which quantum mechanics differs greatly from classical physics is with regards to the statistics of identical particles.  It is well known that there are two different types of quantum particle: bosons and fermions.  The behaviour of the two types of particle is very different.  A characteristic feature of bosons is that they tend to bunch together.  A famous example of this within quantum optics is the Hong-Ou-Mandel (HOM) effect \cite{hom1, hom2}.  This occurs when a photon enters each of the two inputs of a balanced (or 50-50) beam splitter.  In the ideal case we find that the photons have a zero probability of being measured exiting both outputs of the beam splitter.  Instead, we find both photons in one of the two outputs.  One can understand this as being the result of destructive interference between the two possible paths that the photons must take if they are to be found in different outputs \cite{rarity}.  %This is illustrated in figure 1.

The fact that photons show HOM interference is a manifestation of the bosonic nature of photons.  The phenomenon should, in principle, be observed in other bosonic systems.  An interesting feature of bosons is that a system composed of several bosons is itself a boson.  These composite bosons will also obey Bose-Einstein statistics.  In principle, one should be able to demonstrate HOM interference with composite bosons, provided one can find an analogue of an optical beam splitter.

A beam splitter is an example of an optical coupler.  These are devices that allow two modes to interact and be transformed into two different modes.  The operation of an optical coupler can be described by suitable Hamiltonian.  For example, the Hamiltonian $\hat H=\hat a^{\dagger}\hat b+\hat b^{\dagger}\hat a$ can be used to describe a beam splitter, where $\hat a$ and $\hat b$ are the annihilation operators associated with the two optical modes \cite{hamref}.  We work throughout with a dimensionless Hamiltonian, so that our unit of time is defined in terms of the physical coupling constant.  An important class of optical couplers are those that rely on nonlinear optical process.  The reason for the importance of nonlinear optical couplers is that they have been shown to exhibit a number of novel physical phenomena such as second harmonic generation \cite{nonlinear1}, generation of squeezed states \cite{nonlinear2} and collapses and revivals of photon number oscillations \cite{nonlinear3}.  We will investigate a nonlinear Hamiltonian of the form 
\begin{equation}
\label{ham}
\hat H=(\hat a^{\dagger})^2\hat b^2+(\hat b^{\dagger})^2\hat a^2,
\end{equation}
where $\hat a$, $\hat b$ are bosonic annihilation operators that satisfy $[\hat a,\hat b]=0$.%$[\hat a,\hat a^{\dagger}]=\hat 1$ and $[\hat b,\hat b^{\dagger}]=\hat 1$.  
We shall show that the Hamiltonian (\ref{ham}) gives rise to novel dynamics such as HOM interference between two pairs of bosons.  One can think of the Hamiltonian (\ref{ham}) as acting like a beam splitter Hamiltonian, but with photons replaced by pairs of bosons.

The outline of the paper is as follows.  In section \ref{secII} we will investigate the dynamics of the Hamiltonian (\ref{ham}).  In particular we will show that it exhibits HOM interference for composite bosons.   The results will be interpreted in terms of a simple interference picture in section \ref{secIII}.  We will then consider generalizations of the Hamiltonian (\ref{ham}) in section \ref{secIV}.  The case of interference between two composite bosons, each composed of $n$ single bosons, is considered together with more general situations.  A simple example of how the theory can be implemented will be given in section \ref{secV}.  Finally, we will discuss the results in section \ref{concl}

\section{A `beam splitter' for composite bosons}
\label{secII}
Let us investigate the dynamics induced by the Hamiltonian (\ref{ham}).  It can easily be verified that the Hamiltonian, (\ref{ham}), commutes with $\hat a^{\dagger}\hat a+\hat b^{\dagger}\hat b$, hence the total number of bosons will be conserved.  We can, therefore, decompose the Hilbert space into invariant subspaces that have a fixed number of bosons.  By examining the form of (\ref{ham}) we see that the state $|m,n\rangle$ will be coupled with the states $|m-2,n+2\rangle$ and $|m+2,n-2\rangle$.  This implies that the subspaces with fixed boson number can be decomposed further into smaller invariant subspaces that have spanning sets of the form $\{|m,n\rangle,|m\pm 2,n\mp 2\rangle,|m\pm 4,n\mp 4\rangle,...\}$.  

The states $|1,0\rangle$, $|0,1\rangle$ and $|1,1\rangle$ are all eigenvectors of (\ref{ham}) that correspond to the eigenvalue zero.  The dynamics of these states is trivial.  A more interesting situation to consider is when we have two bosons in one mode of the field, e.g. $|2,0\rangle$.  It is clear that the Hamiltonian (\ref{ham}) will couple the states $|2,0\rangle$ and $|0,2\rangle$.  The dynamics that one observes is simply Rabi oscillations between the states $|2,0\rangle$ and $|0,2\rangle$.  %with a Rabi frequency of $2$.  
For a suitably chosen interaction time, we can transform the state $|2,0\rangle$ into $(|0,2\rangle-i|2,0\rangle)/\sqrt{2}$.  This is reminiscent of the behaviour of a balanced beam splitter with respect to a single photon.  The two bosons can be thought of as a single system, a bi-boson, which is itself a boson, albeit one with twice the energy and half the wavelength of a single boson \cite{jbcy}.

Suppose we have a state of the form $|2,2\rangle$.  The Hamiltonian (\ref{ham}) will couple this state to the states $|4,0\rangle$ and $|0,4\rangle$.  The three states $|4,0\rangle$, $|2,2\rangle$ and $|0,4\rangle$ will span a three dimensional subspace that is invariant under the action of equation (\ref{ham}).  We shall denote this subspace as ${\mathcal H}_3$.  If we diagonalize the Hamiltonian (\ref{ham}) within the subspace ${\mathcal H}_3$,  then we find the eigenvectors
\begin{eqnarray}
\label{evectors}
|\lambda_0\rangle&=&\frac{1}{\sqrt{2}}\left(|4,0\rangle-|0,4\rangle\right),\nonumber\\
|\lambda_{\pm}\rangle&=&\frac{1}{2}\left(|4,0\rangle\pm \sqrt{2}|2,2\rangle+|0,4\rangle\right),
\end{eqnarray}
where $\hat H|\lambda_{\pm}\rangle=\pm 4\sqrt{3}|\lambda_{\pm}\rangle$ and $\hat H|\lambda_0\rangle=0$.  The three states $|4,0\rangle$, $|2,2\rangle$ and $|0,4\rangle$, can be expressed in terms of the new basis, (\ref{evectors}), as follows
\begin{eqnarray}
\label{states}
|4,0\rangle&=&\frac{1}{2}\left(|\lambda_+\rangle+|\lambda_-\rangle+\sqrt{2}|\lambda_0\rangle\right),\nonumber\\
|2,2\rangle&=&\frac{1}{\sqrt{2}}\left(|\lambda_+\rangle-|\lambda_-\rangle\right),\nonumber\\
|0,4\rangle&=&\frac{1}{2}\left(|\lambda_+\rangle+|\lambda_-\rangle-\sqrt{2}|\lambda_0\rangle\right).
\end{eqnarray}
We shall denote the evolution operator as $\hat U(t)$.  Within the subspace ${\mathcal H_3}$, the evolution operator has the form $\hat U(t)=\exp(-i\hat H t)=e^{-i\lambda t}|\lambda_+\rangle\langle\lambda_+|+e^{i\lambda t}|\lambda_-\rangle\langle\lambda_-|+|\lambda_0\rangle\langle\lambda_0|$.

Suppose we have four bosons in the first mode, while the second mode is in the vacuum, i.e. the system is prepared in the state $|4,0\rangle$.  This state will evolve under the influence of (\ref{ham}) to the state $\hat U(t)|4,0\rangle=(\sqrt{2}|\lambda_0\rangle+e^{-i4\sqrt{3}t}|\lambda_+\rangle+e^{i4\sqrt{3}t}|\lambda_-\rangle)/2$.  We see that at time $t=\pi/(8\sqrt{3})$, the state has evolved to 
\begin{equation}
\label{twoph}
\hat U\left(\frac{\pi}{8\sqrt{3}}\right)|4,0\rangle=\frac{1}{2}\left(|4,0\rangle-|0,4\rangle-i\sqrt{2}|2,2\rangle\right). 
\end{equation}
The probability of finding two bosons in each mode, at time $t=\pi/(8\sqrt{3})$, is thus $1/2$.  Similarly, at $t=\pi/(8\sqrt{3})$ we find that the probability of finding all four bosons in the first mode is $1/4$, while the probability of finding all the bosons in the second mode is also $1/4$.  From the symmetry of the situation we see that if our state was initially $|0,4\rangle$, then we would obtain the same probabilities for finding the bosons in the given output modes, at time at $t=\pi/(8\sqrt{3})$.  

It is instructive to compare these results to those for a balanced beam splitter with two photons entering one mode, while the other mode is in the vacuum.  It can be shown \cite{loudon} that the probability of obtaining a photon in each output mode is $1/2$, while the probability of having both photons exit in a particular output mode is $1/4$.  The action of the Hamiltonian (\ref{ham}), with respect to pairs of bosons, strongly resembles that of a beam splitter with respect to single photons.  This agreement suggests that the Hamiltonian acts like a balanced beam splitter with respect to pairs of bosons.

Suppose the system is initially prepared with two bosons in each mode of the field, i.e. the state is $|2,2\rangle$.  We thus have one composite boson in each mode.  The Hamiltonian (\ref{ham}) will cause the state to evolve to $\hat U(t)|2,2\rangle=(e^{-i4\sqrt{3}t}|\lambda_+\rangle-e^{i4\sqrt{3}t}|\lambda_-\rangle)/\sqrt{2}$.  The probabilities for detecting the bosons in a particular mode, at time $t$, can be calculated using the previous equation together with equation (\ref{states}).  A simple calculation yields the results
\begin{eqnarray}
\label{p22}
P_{22}(t)&=&|\langle2,2|\hat U(t)|2,2\rangle|^2=\cos^2(4\sqrt{3}t),\\
\label{p40}
P_{40}(t)&=&|\langle4, 0|\hat U(t)|2,2\rangle|^2=\frac{1}{2}\sin^2(4\sqrt{3}t),\\
\label{p04}
P_{04}(t)&=&|\langle0,4|\hat U(t)|2,2\rangle|^2=\frac{1}{2}\sin^2(4\sqrt{3}t).
\end{eqnarray} 
If we choose the interaction time to be $t=\pi/(8\sqrt{3})$, then equation (\ref{p22}) implies that $P_{22}=0$, while $P_{40}=P_{04}=1/2$.  The probability for detecting bosons in both output modes is thus zero.  It can easily be shown that the output state, at $t=\pi/(8\sqrt{3})$, is just 
\begin{equation}
\label{outputhom}
\hat U\left(\frac{\pi}{8\sqrt{3}}\right)|2,2\rangle=\frac{-i}{\sqrt{2}}\left(|4,0\rangle+|0,4\rangle\right).
\end{equation}
The term corresponding to two bosons in each output is, as expected, missing from equation (\ref{outputhom}).   This is analogous to the HOM effect, where we have a photon entering each input of a balance beam splitter, but where we cannot detect photons in both output modes.  We should note that the states given in equations (\ref{outputhom}) and (\ref{twoph}) are both generated using the same interaction time.  %For the sake of definiteness we can think of the bosons as photons.  
For the fixed interaction time, $t=\pi/(8\sqrt{3})$, the action of the Hamiltonian (\ref{ham}), with respect to pairs of bosons, is identical to the behaviour of a balanced beam splitter with respect to single photons.  One can thus think of equation (\ref{ham}) as describing the action of a coupler that treats pairs of bosons as a single system, a bi-boson.  Furthermore, this coupler will act as a balanced beam splitter with respect to bi-bosons.

\section{Interference of composite bosons}
\label{secIII}
In the previous section we showed that the nonlinear Hamiltonian (\ref{ham}), gives rise to dynamics that are analogous to those observed for photons in the HOM effect.  These results can be interpreted in terms of a simple interference picture.  This interpretation will help us to understand the observed multi-boson dynamics.  In particular, it will become clear that the zero probability of obtaining bosons in each mode, i.e. $P_{22}=0$, is a manifestation of HOM interference between two composite bosons.

\begin{figure}[h]
\center{\includegraphics[width=8cm,height=!]
{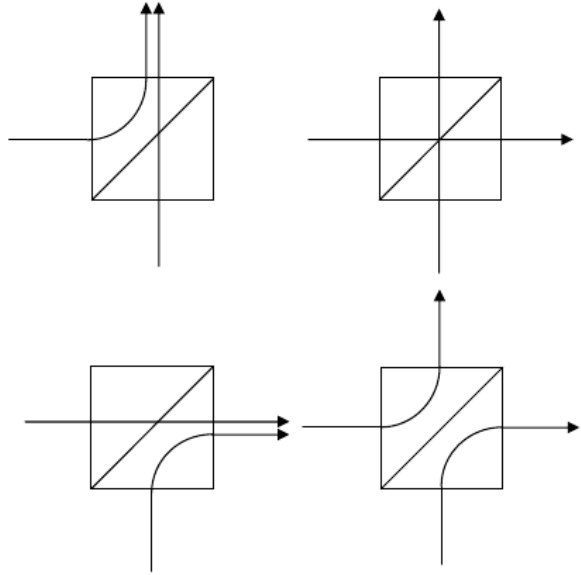}}
\caption{A diagram showing the four paths the the two photons can take through the balanced beam splitter.}
\label{fig1}
\end{figure}

\begin{figure}[!h]
\center{\includegraphics[width=8cm,height=!]
{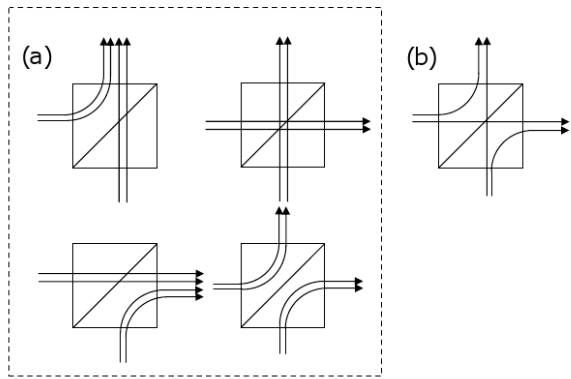}}
\caption{A diagram showing some of the paths the the four bosons can take through the coupler.  Part (a) shows the paths that are consistent with the two bosons being treated as a single composite boson.  Part(b) shows one of the new paths that can occur if the two bosons can travel through the coupler independently.}
\label{fig2}
\end{figure}

To gain insight into the HOM effect we will first look at the familiar case where two photons enter a balanced beam splitter through different inputs.  Each photon can be transmitted or reflected, hence there are four possible paths that the photons can take; this is shown in figure 1.  It can be seen that there is only one path that will lead to both photons being in a particular output.  In contrast, there are two different paths that will result in one photon being in each output mode.  From figure 1 we see that these two different paths correspond to either both photons being transmitted or both photons being reflected.  When a beam of light is reflected it undergoes a phase shift of $\pi/2$ relative to one that was transmitted.  We thus expect that the path corresponding to both photons being reflected will have a phase shift of $\pi$ relative to the case when both photons where transmitted.  This would result in the two paths interfering destructively.  The probability of finding a photon in each output would thus be zero, which is precisely what is observed in the HOM effect.  

The interference picture can be used to interpret the situation where we have bi-bosons entering each input of a coupler.  To allow a simple comparison to the standard HOM effect we will refer to the bosons as photons in the following discussion.  Nevertheless, the results we have obtained apply not only to photons but to any system of bosons.  Suppose we have an optical coupler that has two photons entering each input.  We shall assume that the optical coupler causes the photons to interact in the manner described by (\ref{ham}).  Let us further suppose that the coupler treats pairs of photons (or bi-photons) as a single system.  In this case we will again have four different possible paths for the bi-photons.  Only one path will correspond to finding all the photons in a particular output.  There are, however, two different paths that will result in two photons being in each output of the coupler, see figure 2 a.  If we think of the bi-photons as a single system, then the reflection of a bi-photon will result in a phase difference of $\pi/2$ relative to a bi-photon that was transmitted.  The path where both bi-photons are reflected will thus have a phase difference of $\pi$ relative to the path where both bi-photons are transmitted.  These two paths will destructively interfere in the same manner as we described for the single photon case.  We thus find that the probability of finding two photons in each output will be zero.  We can see that this effect is just two boson interference, where the two bosons are composed of two photons.

To emphasize that the behaviour of section \ref{secII} really is interference of two composite bosons, we will discuss the dynamics that occur if the two bosons are not treated as a single composite system.  An obvious question to ask is: how many paths will result in us obtaining two bosons in each output of the coupler?  To simplify the discussion we will again take the bosons to be photons.  In figure 2 it is shown that there are now three paths that lead to two photons exiting each output\footnote{Actually there are four paths, however, two of these paths are identical.  The indistinguishability of quantum particles means that these two paths should not be viewed as a separate paths.}.  The first two paths correspond to all the photons being either transmitted or reflected.  This is same as for the previous situation, where the photons where treated as a single system.  The new path that we obtain occurs due the fact that the photons in each input can behave differently from each other.  It is thus possible for one photon to be reflected while the other is transmitted.  This will lead to an extra path that cannot be cancelled out.  Furthermore, the two paths that correspond to all the photons being reflected and all the photons being transmitted, will no longer destructively interfere with each other.  The reason for this is that each reflected photon has a $\pi/2$ phase shift, relative to the photons that where transmitted.  The net phase difference between the two paths is thus $2\pi$, which means the two paths will interfere constructively.  We will thus obtain a non-zero probability for detecting two photons in each output mode.  

From the above discussion the following can be concluded.  The Hamiltonian (\ref{ham}) treats pairs of bosons as a single system, a bi-boson.  The effects we analyzed are thus examples of two boson interference.  In particular, the fact that we cannot detect bosons in both outputs, when we have two bosons entering each input, is the result of HOM interference between  two composite bosons.

\section{Interference of general composite bosons}
\label{secIV}
We have shown that the Hamiltonian (\ref{ham}) leads to HOM interference between two composite bosons, which are composed of two other bosons.  An obvious question to ask is whether we can observe similar effects between composite bosons that are composed of $n$ bosons each.  The fact that a system composed of $n$ bosons is also a boson suggests that we should be able to observe HOM interfere, provided we can find a suitable coupler.  The desired action of the coupler is that it should treat a mode with $n$ bosons as containing a single composite boson.  The coupler should then act as a beam splitter with respect to these composite bosons.

Consider the following Hamiltonian
\begin{equation}
\label{nham}
\hat H=(\hat a^{\dagger})^n\hat b^n+(\hat b^{\dagger})^n\hat a^n,
\end{equation}
where $n$ is a positive integer, and $\hat a$ and $\hat b$ are mutually commuting annihilation operators that satisfy bosonic commutator relations.  It can be shown that the total number of bosons is conserved under the action of (\ref{nham}).  The Hamiltonian will thus couple the state with $n$ bosons in each mode,  $|n,n\rangle$, to the states $|2n,0\rangle$ and $|0,2n\rangle$.  These three states will thus span a three dimensional subspace, which is invariant under the action of (\ref{nham}).  We will concern ourselves with diagonalizing (\ref{nham}) within this invariant subspace.  The eigenvectors of $\hat H$ are $|\lambda'_0\rangle=(|2n,0\rangle-|0,2n\rangle)/\sqrt{2}$ and $|\lambda'_{\pm}\rangle=(|2n,0\rangle\pm \sqrt{2}|n,n\rangle+|0,2n\rangle)/2$, where $\hat H|\lambda'_0\rangle=0$ and $\hat H|\lambda'_{\pm}\rangle=\pm\sqrt{(2n)!}|\lambda'_{\pm}\rangle$.  A straightforward calculation shows that at time $t=\pi/(2\sqrt{(2n)!})$, the probability of obtaining $n$ bosons in each output mode, when $n$ bosons enter each input, is zero.  The Hamiltonian (\ref{nham}) thus gives rise to HOM interference between two composite bosons that are each composed of $n$ bosons.

Thus far we have investigated interference between two composite bosons that are each composed of $n$ bosons.  An interesting question to consider is whether or not one can observe HOM interference between two composite bosons that are composed of different numbers of bosons.  For example, we may wish to have a field containing two photons interacting with another field that contains only one photon.  In order to answer this question we introduce the following Hamiltonian 
\begin{equation}
\label{gham}
\hat H=(\hat a^{\dagger})^n\hat b^k+(\hat b^{\dagger})^k\hat a^n,\;\;n,k\in{\mathbb N}:n,k>0,
\end{equation}
where $\hat a$ and $\hat b$ are bosonic annihilation operators that act on the first and second modes respectively.  Hamiltonians of the form of (\ref{gham}) have been studied before \cite{goce, igor} and can be used in models of nonlinear couplers \cite{nc1, nc2, nc3}.  It can be verified that $\hat M=k\hat a^{\dagger}\hat a+n\hat b^{\dagger}\hat b$ commutes with (\ref{gham}).   The systems Hilbert space can thus be decompose into subspaces with a fixed eigenvalue for $\hat M$.  For example, the states $|n,k\rangle$, $|2n,0\rangle$ and $|0,2k\rangle$ are eigenvectors of $\hat M$ that correspond to the eigenvalue $2nk$.

Suppose we prepare a system with $n$ bosons in the first mode and $k$ bosons in the second, i.e. it is prepared in the state $|n,k\rangle$.  The general Hamiltonian (\ref{gham}) will couple the state $|n,k\rangle$ to the states $|2n,0\rangle$ and $|0,2k\rangle$.  These three states will span a three dimensional subspace that is invariant under the action of (\ref{gham}).  We will confine ourselves to this three dimensional subspace.  The structure of the problem is clarified somewhat if we represent (\ref{gham}) in matrix form.  We set $|2n,0\rangle=(1,0,0)^T$, $|n,k\rangle=(0,1,0)^T$ and $|0,2k\rangle=(0,0,1)^T$, which yields the following form for (\ref{gham})
\begin{equation}
\label{matr} 
H=\left[ 
\begin{matrix}
0 & x & 0\\
x & 0 & y\\
0 & y & 0
\end{matrix}
\right],
\end{equation}
where $x$ and $y$ are
\begin{equation}
\label{xy}
x=\sqrt{\frac{(2n)!k!}{n!}},\;\;y=\sqrt{\frac{(2k)!n!}{k!}}.
\end{equation}
One can easily check that the eigenvectors of (\ref{matr}) are
\begin{eqnarray}
\label{eigenv}
|\psi_0\rangle&=&\frac{1}{\lambda}\left(y|2n,0\rangle-x|0,2k\rangle\right),\nonumber\\
|\psi_{\pm}\rangle&=&\frac{1}{\sqrt{2}\lambda}\left(x|2n,0\rangle\pm\lambda|n,k\rangle+y|0,2k\rangle\right),
\end{eqnarray} 
where $\lambda=\sqrt{x^2+y^2}$.  It can also be seen that $\hat H|\psi_0\rangle=0$, $\hat H|\psi_{\pm}\rangle=\pm \lambda|\psi_{\pm}\rangle$.  We can expand the three vectors $|2n,0\rangle$, $|n,k\rangle$ and $|0,2k\rangle$ in terms of the eigenvectors (\ref{eigenv})
\begin{eqnarray}
|2n,0\rangle&=&\frac{1}{\lambda}\left(y|\psi_0\rangle+\frac{x}{\sqrt{2}}|\psi_+\rangle+\frac{x}{\sqrt{2}}|\psi_-\rangle\right),\nonumber\\
|n,k\rangle&=&\frac{1}{\sqrt{2}}\left(|\psi_+\rangle-|\psi_-\rangle\right),\nonumber\\
|0,2k\rangle&=&\frac{1}{\lambda}\left(\frac{y}{\sqrt{2}}|\psi_+\rangle+\frac{y}{\sqrt{2}}|\psi_-\rangle-x|\psi_0\rangle\right).
\end{eqnarray}

The system will evolve from the initial state $|n,k\rangle$ to the state $\hat U(t)|n,k\rangle$ at time $t$.  If we place detectors at each output, then we will obtain one of the following three outcomes.  We detect $2n$ bosons in the first mode, we detect $2k$ bosons in the second mode, or we detect $n$ bosons in the first mode and $k$ bosons in the second mode.  The probabilities of these three outcomes can be easily calculated and are found to be
\begin{eqnarray}
\label{t1}
P_{2n,0}(t)&=&|\langle 2n,0|\hat U(t)|n,k\rangle|^2=\frac{x^2}{x^2+y^2}\sin^2(\lambda t)\\
\label{t2}
P_{0,2k}&=&|\langle 0,2k|\hat U(t)|n,k\rangle|^2=\frac{y^2}{x^2+y^2}\sin^2(\lambda t)\\
\label{t3}
P_{n,k}&=&|\langle n,k|\hat U(t)|n,k\rangle|^2=\cos^2(\lambda t).
\end{eqnarray}
If we choose our interaction time to be $t=\pi/(2\lambda)$, then the probability of obtaining bosons in both output modes, given in equation (\ref{t3}), will be zero.  This is again an example of HOM interference and can be explained using the same interference picture that we described in section \ref{secIII}.  One difference between the current situation and those discussed earlier is that now the probabilities for finding all the bosons in a given output are different for each output.  Using equations (\ref{t1}) and (\ref{t2}), we find that at $t=\pi/(2\lambda)$, the probabilities for finding all the bosons in a single output are
\begin{equation}
\label{factorial}
P_{2n,0}=N\frac{(2n)!k!}{n!}\;\;\text{and}\;\;P_{0,2k}=N\frac{(2k)!n!}{k!},
\end{equation}
where $N=1/(x^2+y^2)$.  If we ignore the normalization terms in equation (\ref{factorial}), then we see that the probabilities are combinatorial terms that describe the number of different ways of distributing $n$ and $k$ indistinguishable particles between two modes \cite{dirac}.

We have diagonalized (\ref{gham}) within the subspace spanned by the vectors $|2n,0\rangle$, $|n,k\rangle$ and $|0,2k\rangle$.  We could also look at different invariant subspaces.  For instance the three dimensional subspace spanned by $|2n+s,r\rangle$, $|n+s,k+r\rangle$ and $|s,2k+r\rangle$ will also be an invariant subspace provided $r$ and $s$ are integers within the ranges $0\le s<n$ and $0\le r<k$.  If we set $|2n+s,r\rangle=(1,0,0)^{T}$, $|n+s,k+r\rangle=(0,1,0)^{T}$ and $|s,2k+r\rangle=(0,0,1)^{T}$, then our Hamiltonian, (\ref{gham}), will have the same form as (\ref{matr}), but where $x$ and $y$ have the values
\begin{equation}
x=\sqrt{\frac{(2n+s)!(k+r)!}{(n+s)!r!}}\;\;\text{and}\;\;y=\sqrt{\frac{(2k+r)!(n+s)!}{(n+r)!s!}}.
\end{equation}
The dynamics within this subspace is formally the same as in the previous case.  In particular the probability of finding $n+s$ bosons in the first mode and $k+r$ in the second mode, has the same form as (\ref{t3}).  A straightforward calculation shows that
\begin{equation}
\label{p3}
P_{n+s,k+r}(t)=|\langle n+s,k+r|\hat U(t)|n+s,k+r\rangle|^2=\cos^2(\lambda t).
\end{equation}
If we choose the interaction time to be $t=\pi/(2\lambda)$, then we find that $P_{n+s,k+r}(t)=0$.  We can interpret this effect as an example of HOM interference between two composite bosons that are composed of $n$ and $k$ bosons respectively.  

All of the examples that we have considered have consisted of two modes.  We can, however, develop an analogous theory for systems with more than two modes.  We will demonstrate this be investigating a system that is composed of three modes.  Suppose we have three modes that are coupled by a Hamiltonian of the following form 
\begin{equation}
\label{3h}
\hat H=(\hat c^{\dagger})^n\hat a^k\hat b^k+(\hat a^{\dagger})^k(\hat b^{\dagger})^k\hat c^n,
\end{equation}
where $\hat a$, $\hat b$ and $\hat c$ are bosonic annihilation operators that act on the first, second and third system respectively.  This Hamiltonian will couple the states $|n,k,k\rangle$, $|2n,0,0\rangle$ and $|0,2k,2k\rangle$.  If we set $|2n,0,0\rangle=(1,0,0)^T$, $|0,2k,2k\rangle=(0,1,0)^T$ and $|0,2k,2k\rangle=(0,0,1)^T$, then (\ref{3h}) will be a matrix with the same form as (\ref{matr}), but where $x$ and $y$ have the form
\begin{equation}
x=k!\sqrt{\frac{(2n)!}{n!}},\;\;y=\sqrt{n!}\frac{(2n)!}{k!}.
\end{equation}
The dynamics of this system will thus be formally the same as for the general Hamiltonian (\ref{gham}), within the three dimensional subspace.  In particular we can observe HOM interference, i.e. we can choose the interaction time so that a state $|n,k,k\rangle$ will be transformed to a superposition of the states $|2n,0,0\rangle$ and $|0,2k,2k\rangle$.

\section{Example of Hong-Ou-Mandel interference in the Dicke model}
\label{secV}
In the previous sections we have shown how assemblies of bosons will interact as a single composite system, under the action of a suitable Hamiltonian.  We have not, however, discussed how one could realize the theory in an experimentally feasible context. In this section we will give a concrete example.  

%Consider the following situation.  
Suppose we have two atoms in a cavity that are interacting with a single mode of the electromagnetic field within the cavity.  For simplicity we shall assume that the cavity mode is resonant with a particular transition between two energy levels of the atoms and consider only this transition for each atom.  The atoms can be thought of as having only two energy levels, the ground state, $|g\rangle$ and the excited state $|e\rangle$.  This is an example of a two atom Dicke model \cite{dicke}.  We will consider only the symmetric states of the two atoms.  We can thus represent the atomic state in the form $|\widetilde{m}\rangle$, where $m$ equals the number of atoms in the excited state.  For instance, $|\widetilde{0}\rangle=|g,g\rangle$, $|\widetilde{1}\rangle=(|e,g\rangle+|g,e\rangle)/\sqrt{2}$ and $|\widetilde{2}\rangle=|e,e\rangle$.  We can define the individual atomic dipole operators as follows: $\hat\sigma^{(j)}_{+}$ will raise the $j$-th atom from the ground state to the excited state, and $\hat\sigma^{(j)}_-$ will lower the $j$-th atom from the excited state to the ground state.  We also define $\hat\sigma_z^{(j)}=2\hat\sigma^{(j)}_+\hat\sigma^{(j)}_--1$.  The total atomic dipole operators, $\hat S_{z}$, $\hat S_{\pm}$ will be taken to be the sum of the individual atomic dipole operators, hence $\hat S_{z,\pm}=\hat\sigma^{(1)}_{z,\pm}+\hat\sigma^{(2)}_{z,\pm}$.  One can verify that 
\begin{eqnarray}
\hat S_{+}|\widetilde{m}\rangle&=&\sqrt{2}|\widetilde{m+1}\rangle,\nonumber\\
\hat S_-|\widetilde{m+1}\rangle&=&\sqrt{2}|\widetilde{m}\rangle,
\end{eqnarray}
where $m=0,1$.

A useful way of describing this system is to make use of the Schwinger representation for angular momentum \cite{schwing}.  In this approach we introduce two fictitious bosonic modes $\hat b_1$ and $\hat b_2$.  We then make the following substitution for the atomic dipole operators 
\begin{eqnarray}
\hat S_+&=&\hat b_1\hat b_2^{\dagger},\nonumber\\
\hat S_-&=&\hat b_1^{\dagger}\hat b_2\nonumber\\
\hat S_z&=&(\hat b_2^{\dagger}\hat b_2-\hat b_1^{\dagger}\hat b_1)/2.
\end{eqnarray}
The atomic states can be represented by Fock states $|j,j'\rangle$, where $|\widetilde{m}\rangle=|2-m,m\rangle$.  The Fock state $|1,1\rangle$ is thus equivalent to the symmetric state $(|e,g\rangle+|g,e\rangle)/\sqrt{2}$.  The Hamiltonian that describes the interaction between the two atoms and the field mode will have the form \cite{opo2,jex, perina}
\begin{equation}
\label{dh}
\hat H_I=\hat a\hat b_1\hat b_2^{\dagger}+\hat a^{\dagger}\hat b^{\dagger}_1\hat b_2,
\end{equation}
where $\hat a$ is the annihilation operator for the cavity mode.  We can see that the Hamiltonian (\ref{dh}) has the same form as (\ref{3h}), where $n=k=1$.  The total state of the cavity field and the atoms can be expressed in the form $|N,j,j'\rangle$, where $N$ is the number of photons in the cavity mode and $|j,j'\rangle$ is the Fock state that represents the state of the atom in the Schwinger representation.  

The cavity field will be prepared so that it contains a single photon, while the atom pair is initially prepared in the symmetric state $|1,1\rangle$.  The initial state of the full system is thus $|1,1,1\rangle$.  After the atoms and cavity mode have interacted for a time $t=\pi/(2\lambda)$, the state of the system will have been transformed to
\begin{equation}
\label{dickehom}
\hat U\left(\frac{\pi}{2\lambda}\right)|1,1,1\rangle=\frac{-i}{\sqrt{2}}\left(|2,2,0\rangle+|0,0,2\rangle\right).
\end{equation}
The state (\ref{dickehom}) does not contain the term $|1,1,1\rangle$.  We can view this as HOM interference between two composite bosons, where the first composite boson consists of a photon and an atom, while the second boson is composed of a single atom.

\section{Conclusion}
\label{concl}
We have investigated a special type of multi-boson interference.  A Hamiltonian, (\ref{ham}), was presented that treats two bosons as a single composite boson.  We have showed that the Hamiltonian allows two composite bi-bosons to display HOM interference.  The interpretation of this effect was corroborated by a simple interference picture.  In particular, we showed that the observed phenomenon was a direct consequence of the two bosons acting as a single composite boson.  If the two bosons had acted independently, then we would have observed bosons in both outputs.   

The task of engineering a coupler that causes the bosons to interact in the manner described by (\ref{ham}) is non-trivial.  In practical situations we could have higher order terms appearing in the Hamiltonian.  One might expect that these terms will always lead to the interference being washed out.  This, however, is not always true.  As an example, consider adding the higher order terms  $\lambda[\hat a^4(\hat b^{\dagger})^4+(\hat a^{\dagger})^4\hat b^4]$ to the Hamiltonian (\ref{ham}). In this situation the states $|4,0\rangle$, $|2,2\rangle$ and $|0,4\rangle$ are still coupled.  Furthermore, it can be shown that we still observe HOM interference, but where now the interaction time needed to observe this will have changed. 

We have shown that within the 2 and 4 boson subspaces, the Hamiltonian (\ref{ham}), acts like a beam splitter with respect to bi-bosons.  This will not hold within higher dimensional subspaces.  It can be shown that the dynamics of $N$ bi-bosons, $N>2$, is not the same as the dynamics of $N$ photons interacting via a beam splitter Hamiltonian.  This is because the matrix elements of two Hamiltonians are different within the corresponding invariant subspaces. 

An assembly of $N$ bosons can be viewed as a single composite system, which is also a boson.  This observation means that it is possible to observe two boson interference between two general composite bosons, provided one can find a suitable interaction.  We presented a general Hamiltonian, (\ref{gham}), that would allow one to observe interference between composite bosons that where composed of $n$ and $k$ bosons.  We were able to observe HOM interference between these two composite bosons.  This general type of two boson interference exhibited interesting effects such as the following.  For $n=k$, the probability of obtaining all the bosons in the first output was the same as the probability of obtaining all the bosons in the second output.  When $n\ne k$, we find that these two outcomes occur with different probabilities.  %To explain why this was the case consider the situation where we observed $$2n$ The reason for this was that when we observed $2n$ photons in one output then we have had to change the $k$ photons in the second input into $n$ photons.  The fact that 
The reason for this was that we observe different numbers of bosons in each output, either $2n$ or $2k$.  The probabilities for observing $2n$ or $2k$ bosons in a given output are, up to a normalization constant, just combinatorial terms that give the number of ways of removing $n$ or $k$ indistinguishable bosons and then redistributing $k$ or $n$ new bosons with the remaining bosons in the given output.

Finally, we showed how the two atom Dicke model could give rise to dynamics that could be interpreted as HOM interference between two composite bosons.  In this case the interference was between an atom and photon pair and a single atom.  In principle one could observe interference between two composite bosons in other systems, such as Bose-Einstein condensates, the $N$ atom Dicke model or between phonons in trapped ions.  The implementation in terms of phonons is of particular interest as the vibrational dynamics of trapped ions operating within the resolved-sideband regime, can be described by a nonlinear Hamiltonian \cite{vogel, leib}.

\section*{Acknowledgements}
We acknowledge financial support from the Doppler Institute and from grants MSM6840770039 and MSMT
LC06002 of the Czech Republic. S.M.B. thanks the Royal Society and the Wolfson Foundation for financial support.

\end{document}